\documentclass[journal]{IEEEtran}
\IEEEoverridecommandlockouts

\usepackage{xfrac}
\usepackage{amsmath}
\usepackage{multirow,etoolbox}
\usepackage{color}
\usepackage{amssymb}
\usepackage{cite}
\usepackage{amssymb,amsmath,mathtools}

\usepackage{diagbox}

\usepackage{subcaption}
\usepackage[utf8]{inputenc}

\usepackage[subtle]{savetrees}

\usepackage{stfloats}
\usepackage{relsize}

\makeatletter
\patchcmd{\@maketitle}
  {\addvspace{0.5\baselineskip}\egroup}
  {\addvspace{-1\baselineskip}\egroup}
  {}
  {}
\makeatother

\begin{document}
\bstctlcite{IEEEexample:BSTcontrol}

\title{\makebox[\linewidth]{\parbox{\dimexpr\textwidth+1.5cm\relax}{\centering Error Analysis of Cooperative NOMA with Practical Constraints: Hardware-Impairment, Imperfect SIC and CSI}}}

\author{Beddiaf Safia, Khelil Abdellatif, Faical Khennoufa, Ferdi Kara,~\IEEEmembership{Senior Member,~IEEE,}  Hakan Kaya, Xingwang Li,~\IEEEmembership{Senior Member,~IEEE,} Khaled Rabie,~\IEEEmembership{Senior Member,~IEEE,} and Halim Yanikomeroglu,~\IEEEmembership{Fellow,~IEEE.} 
\thanks{B. Safia, K. Abdellatif and F. Khennoufa are with Department of Electrical Engineering, Echahid Hamma Lakhdar University, El-Oued, Algeria, email:\{beddiaf-safia, abdellatif-khelil, khennoufa-faical \}@univ-eloued.dz.} 
\thanks{F. Kara is with the Department of Computer Engineering, Zonguldak Bulent Ecevit University, Zonguldak, Turkey, 67100,  e-mail: f.kara@beun.edu.tr.}
\thanks{H. Kaya is with the Electical and Electronics Engineering, Zonguldak Bulent Ecevit University, Zonguldak, Turkey, 67100,  e-mail: hakan.kaya@beun.edu.tr.}
\thanks{X. Li is with School of Physics and Electronic Information Engineering, Henan Polytechnic University, Jiaozuo, China, e-mail: lixingwang@hpu.edu.cn.}
\thanks{K. Rabie is with the Department of Engineering, Manchester Metropolitan University, Manchester M1 5GD, U.K. e-mail: k.rabie@mmu.ac.uk.}
\thanks{H. Yanikomeroglu is with the Department of Systems and Computer Engineering, Carleton University, Ottawa, K1S 5B6, ON, Canada, e-mail: halim@sce.carleton.ca.}
        }
\maketitle
\begin{abstract}
Non-orthogonal multiple access (NOMA) has been a strong candidate to support massive connectivity in future wireless networks. In this regard, its implementation into cooperative relaying, named cooperative-NOMA (CNOMA), has received tremendous attention by researchers. However, most of the existing CNOMA studies have failed to address practical constraints since they assume ideal conditions. Particularly, error performance of CNOMA schemes with imperfections has not been investigated, yet. In this letter, we provide an analytical framework for error performance of CNOMA schemes under practical assumptions where we take into account  imperfect successive interference canceler (SIC), imperfect channel estimation (ICSI), and hardware impairments (HWI) at the transceivers. We derive bit error rate (BER) expressions in CNOMA schemes whether the direct links between source and users exist or not which is, to the best of the authors' knowledge, the first study in the open literature. For comparisons, we also provide BER expression for downlink NOMA with practical constraints which has also not been given in literature, yet. The theoretical BER expressions are validated with computer simulations where the perfect-match is observed. Finally, we discuss the effects of the system parameters (e.g., power allocation, HWI level) on the performance of CNOMA schemes to reveal fruitful insights for the society.

\end{abstract}
\IEEEpeerreviewmaketitle
\section{Introduction}
Non-orthogonal multiple access (NOMA) is a promising solution for satisfying the demands for spectrum efficiency and network density in the next generations of wireless access. Therefore, the integration of the
NOMA with cooperative relaying (CNOMA) networks has attracted recent attention to improve transmission reliability and extend
network coverage and increase spectral efficiency \cite{zeng2020cooperative}. 

In CNOMA schemes, there are two possible configurations depending on whether direct links are available between source and users or not. In the first, the relay node(s) improves the key performance indicators (KPIs), e.g., capacity, outage probability and bit error rate (BER), while in the second, the relay node(s) extends the coverage. In \cite{Liu2018}, the authors investigate the capacity and outage performances for both cases. Nevertheless, almost all literature is devoted to second scenario where the direct links are blocked by some obstacles. On the other hand, most of the existing CNOMA schemes take into account perfect assumptions where practical constraints are disregarded at whole or in part. However, assuming a perfect successive interference canceler (SIC) for NOMA or perfect hardware implementation with no imperfection or ideal channel state information (CSI) in the communication system are not realistic. In practical scenarios, CNOMA schemes suffer from imperfect SIC, CSI errors and also from hardware impairments such as oscillator phase noise, high power amplifier, and in-phase and quadrature-phase imbalance. \cite{smaini2012rf}. Nevertheless, several studies have examined the effects of such imperfections on the performances of CNOMA schemes. An amplify-and-forward (AF) CNOMA system with HWI is analyzed in terms of outage probability and intercept probability over multipath fading channel in \cite{li2019effects}. The authors in \cite{deng2020hardware} investigate the outage probability and ergodic rate performance of the full-duplex (FD) CNOMA with the presence of HWI over Rician fading channels. The authors in \cite{li2019residual} study the impacts of HWI on the NOMA with imperfect CSI (ICSI), where both cooperative and non-cooperative NOMA are analyzed in terms of outage probability, ergodic capacity and energy efficiency. In \cite {arzykulov2021hardware}, the outage probability of the cognitive radio NOMA network is analyzed under HWI, ICSI and imperfect successive interference cancellation (ipSIC). On the other hand, BER performances of FD-CNOMA have been investigated in \cite{hamza2021error} with ipSIC where perfect CSI and no HWI are considered. Then, the BER and outage probability performances of a multi-hop decode-and-forward (DF) CNOMA with ICSI has been investigated in \cite{khennoufa2022bit}. The pairwise error probability (PEP) is derived for the AF-CNOMA by considering HWI with no ICSI. \cite{mohjazi2019error}.

All aforementioned CNOMA studies investigate the performances of CNOMA without direct links (DL) and CNOMA with DL has not been investigated with imperfections in terms of any performance metrics. Besides, most of above papers \cite{zeng2020cooperative, Liu2018, li2019effects, deng2020hardware, li2019residual, arzykulov2021hardware} analyze the effects of one or multiple imperfections only in terms of information-theoretic perspectives (e.g., achievable rate or outage probability) whilst the BER performance (i.e., one of the most important KPIs) is considered in very limited studies \cite{hamza2021error,khennoufa2022bit,mohjazi2019error}. Furthermore, the effects of HWI on the BER performances of NOMA with/without cooperative have not been studied well although its effects on capacity and outage performances have been evaluated in \cite{li2019effects,deng2020hardware,li2019residual,arzykulov2021hardware}. As discussed above, to the best of our knowledge, the BER performance of NOMA schemes (with/without cooperative relaying) has not been investigated with practical constraints (i.e., ipSIC, ICSI and HWI). Besides, the performance of CNOMA with DL has not been revealed with any imperfections in the open literature. 

Motivated by these discussions, in this letter, we provide an analytical framework for the BER performances of NOMA schemes under practical constraints where we consider ipSIC, ICSI and HWI. We derive exact BER expressions for three NOMA schemes: downlink NOMA, CNOMA without DL and CNOMA with DL. We validate our analysis through extensive simulations and discuss the effects of the system parameters (e.g., power allocation or HWI level) on the BER performances of NOMA schemes

The rest of the letter is organized as follows. The CNOMA schemes with/without direct links are introduced in Section II. In Section III, we analyze the exact end-to-end (e2e) BER expressions. The simulation results are presented in Section IV. Finally, Section V concludes the letter.

\section{System Model}
We consider a downlink cooperative NOMA system as presented in Fig. 1. The model consists of a source (S), DF relay and two users: user 1 (u1) and user 2 (u2). The S, relay (R) and users are equipped with a single antenna and the relay works in half duplex (HD) mode. We assume three different scenarios of the considered system which are : $i) \ Downlink \ NOMA:$ S transmits directly the signal to the users, namely  NOMA. $ii) \ Cooperative\ NOMA\ without\ direct\ link:$ S communicates with users with the aid of DF relay without direct link between the S and users, namely CNOMA. $iii)\ Cooperative \ NOMA \ with \ direct \ link:$ S sends the signal to the user through the DF relay and the direct links, where the users combine both received signals in two phases, namely CNOMA-WDL. The communication is completed in a single phase for downlink NOMA whereas it requires two phases in cooperative schemes. The complex flat fading channel coefficient between S-users, R-users and S-R are denoted  as $h_{si} \sim \mathcal{CN}(0,\ \sigma_{si}^2)$, $h_{ri} \sim \mathcal{CN}(0,\ \sigma_{ri}^2)$ and  $h_{sr} \sim \mathcal{CN}(0,\ \sigma_{sr}^2)$ where $i=1,\ 2$, respectively.  $\sigma_{{h}_{si}}^2\ ={d}_{h_{si}}^{-a}$, $\sigma_{{h}_{ri}}^2\ ={d}_{h_{ri}}^{-a}$ and $\sigma_{{h}_{sr}}^2\ ={d}_{h_{sr}}^{-a}$, where ${d}_{h_{si}}$, ${d}_{h_{ri}}$ and ${d}_{h_{sr}}$ are distances between related nodes and $a$ is the path loss exponent. We assume that the ICSI exists for each node. The estimated channel coefficients are given as $\tilde{{h}}_{si}  = {h}_{si}- e $, $\tilde{{h}}_{ri}  = {h}_{ri}- e $ and $\tilde{{h}}_{sr}  = {h}_{sr}- e $, where the channel estimation error is presented as $e \sim \mathcal{CN}(0,\ \sigma_\epsilon^2)$. Since $\tilde{{h}}_{si}$, $\tilde{{h}}_{ri}$, $\tilde{{h}}_{sr}$ and ${e}$ are independent, then it can be modeled as $\sigma _{\tilde{{h}}_{si}}^2\ = \sigma_{{h}_{si}} ^2\ -\sigma_\epsilon^2$, $\sigma_{\tilde{{h}}_{ri}}^2\ = \sigma_{{h}_{ri}} ^2\ -\sigma_\epsilon^2  $ and $\sigma_{\tilde{{h}}_{sr}}^2\ = \sigma_{{h}_{sr}} ^2\ -\sigma_\epsilon^2 $ \cite{afana2018joint}.
\begin{figure}
    \centering
    \includegraphics[width=0.9\columnwidth]{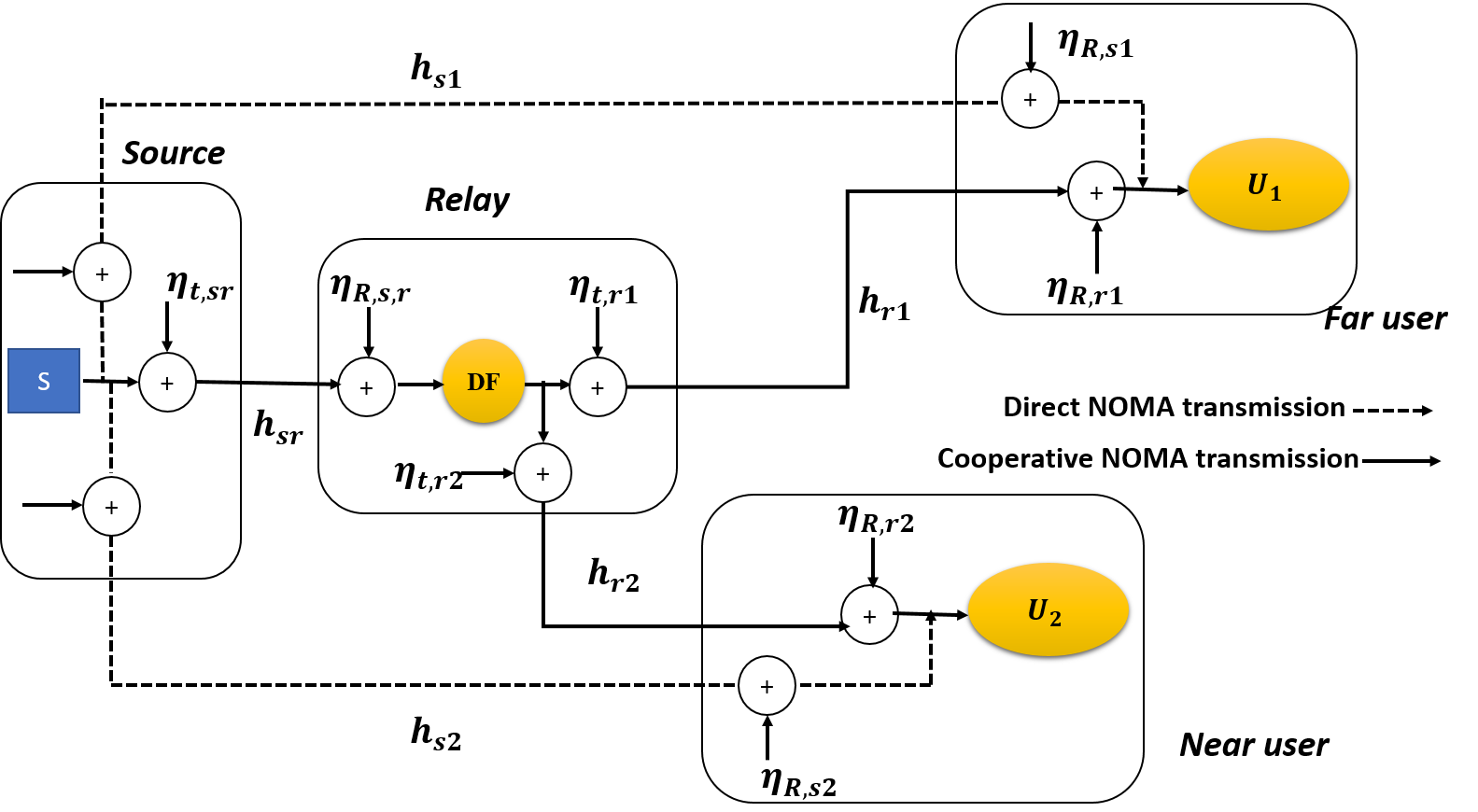}
    \caption{System model CNOMA schemes with HWI.}
    \label{constellations}
\end{figure}
As presented in Fig 1, the S transmits a superimposed coding (SC) signal
\begin{math}
{s}_{sc}= \sqrt{\alpha_1}m_1+\sqrt{\alpha_2}m_2
\end{math} to R and users
with different power coefficients according to channel gain $|{h}_{s1}|^2< |{h}_{s2}|^2$. The received signals at u1, u2 and R (for cooperative schemes) are given as 
\begin{equation}
{y}_{si}= (\tilde{{h}}_{si}\ + {e}) (\sqrt{P_s}{s}_{sc}+\eta_{t,{si}})+\eta_{R,si}+ {n}_{si},
\end{equation}
\begin{equation}
{y}_{sr}= (\tilde{{h}}_{sr}\ + {e}) (\sqrt{P_s}{s}_{sc}+\eta_{t,{sr}})+\eta_{R,sr}+ {n}_{sr},
\end{equation}
where $m_1$, $m_2$ are the message of the u1 and u2, and $\alpha_1$, $\alpha_2$ are the PA coefficients of u1 and u2, respectively. $\alpha_1+\alpha_2=1$. $P_s$  is the transmit power of S. $\eta_{t,si}$, $\eta_{t,sr}$ and $\eta_{R,si}$, $\eta_{R,sr}$  are distortion noises at the transmitter and receiver, respectively, which occur due to the HWI at transceivers  such as oscillator phase noise (PN), high power amplifier (HPA) distortion and in-phase and quadrature-phase imbalance (IQI). We assume  ${n}_{si}={n}_{sr}={n}$ and it is the additive white Gaussian noise (AWGN) which follows $\ {n} \sim \mathcal{CN}(0,\ \frac{N_0}{2})$. The distortion noises are defined as
\begin{equation}
\begin{split}
& \mathbf {\eta}_{t,{si}} \sim \mathcal{CN}(0,\ {k}_{t,{si}}^2 {P}_s),   \  \mathbf {\eta}_{R,{si}} \sim \mathcal{CN}(0,\ {k}_{R,{si}}^2 {P}_s |{h}_{si}|^2), \\&
 \mathbf {\eta}_{t,{sr}} \sim \mathcal{CN}(0,\ {k}_{t,{sr}}^2 {P}_s),   \  \mathbf {\eta}_{R,{sr}} \sim \mathcal{CN}(0,\ {k}_{R,{sr}}^2 {P}_s |{h}_{sr}|^2),  \end{split}
 \end{equation}
where ${k}_{t,{si} }^2$, ${k}_{t,{sr} }^2$ and ${k}_{R,{si}}^2$, ${k}_{R,{sr}}^2$ represent level of impairment at the transmitter and receiver, respectively. Then, (1) can be regarded as 
 \begin{equation}
{y}_{si}= (\tilde{{h}}_{si}\ + {e}) (\sqrt{P_s}{s}_{sc}+\eta_{si}) + {n},
\end{equation}
\begin{equation}
{y}_{sr}= (\tilde{{h}}_{sr}\ + {e}) (\sqrt{P_s}{s}_{sc}+\eta_{sr}) + {n},
\end{equation}
where $\eta_{si}$ and $\eta_{sr}$ are the independent distortion noise terms defined as $ \mathbf {\eta}_{si} \sim \mathcal{CN}(0,\ {k}_{si}^2 {P}_s)$, $ \mathbf {\eta}_{sr} \sim \mathcal{CN}(0,\ {k}_{sr}^2 {P}_s)$, $k_{si}^2$ and $k_{sr}^2$ are the HWI level at the transceiver.
It has been demonstrated in\cite{afana2018joint} that the impact of the transceiver HWI can be characterized by the aggregate level of impairments, $k_{si}^2={k}_{t,{si}}^2 + {k}_{R,{si}}^2$ and $k_{sr}^2 = {k}_{t,{sr}}^2 + {k}_{R,{sr}}^2$.

In the downlink NOMA, the u1 decodes its message directly by maximum-likelihood detection (MLD), the u2 and relay performs an SIC to detect $m_1$ and $m_2$. 

On the other hand, in the cooperative schemes, the R implements an SIC process to obtain both users' symbols. Then, R implements a new SC signal and forwards it by the power of the R to the users in the second phase. Again, R distributes the values of the power coefficient according to channel gain as in S. The received signal at u1 and u2 in the second phase is given as
\begin{equation}
{y}_{ri}= (\tilde{{h}}_{ri}\ + {e})(\sqrt{P_r} {s}_{sc}+\eta_{ri}) + {n},
\end{equation}
where $P_r$  is the relay transmit power and $ \mathbf {\eta}_{ri} \sim \mathcal{CN}(0,\ {k}_{ri}^2 {P}_r)$. 

In CNOMA, the users receive only the second phase signal; hence, they decode their messages according to ${y}_{ri}$ (6) by using MLD or SIC. However, in CNOMA-WDL, the users receive signals in both phases. Thus, the users first implement a maximum ratio combining by using ${y}_{si}$ (1) and ${y}_{ri}$ (6), then, they implement a proper detection scheme (e.g., MLD or SIC).

\section{Performance Analysis}
In this section, we derive the BER expressions of NOMA, CNOMA and CNOMA-WDL for both users over  Rayleigh fading channels. The binary phase-shift keying (BSPK) modulation is used to modulate the message of u1 and u2 at S and R.
The e2e BER expressions in the CNOMA for both users are given as \cite{khennoufa2022bit} by 
\begin{equation}
{P}_{e2e,i}^{(CNOMA)}= {P}_{m_i,sr}
\left(1-{P}_{m_i,ri}\right)+\left(1-{P}_{m_i,sr}\right){P}_{m_i,ri}, 
\end{equation}
where ${P}_{m_i,sr}$ is the BER of the u1 and u2 symbols in relay at first phase, ${P}_{m_i,ri}$ is BER of u1 and u2 in second phase. 

On the other hand, the e2e BER expressions in CNOMA-WDL for both users are given as \cite{kara2018error} by 
\begin{equation}
\begin{split}
&P_{e2e,i}^{(CNOMA-WDL)}=\\
&\frac{1}{2}\sum\limits_{f}g_f [P_{m_i,prop}\times P_{m_i,sr}
 +(1-P_{m_i,sr}) P_{m_i,coop}], \ f=z, \ v,
\end{split}
\end{equation}
where $g_f, \ f=z,v$ is a parameter for u1 and u2, respectively, which is determined according to modulation order. It will be defined in the following related subsections. $P_{m_1,prop}$ is the BER in the presence of error propagation from R to users achieved by the MRC and $P_{m_i,coop}$ is the BER when the symbols of users are detected correctly at the relay and forwarded to the users and then combined with MRC at the users.
\subsection{BER of NOMA for point-to-point (P2P) communication}
In P2P communication, each node (i.e., R, u1 and u2) performs an MLD to detect $m_1$ symbol. Thus, the BER for $m_1$ symbols in the presence of the effects of HWI and ICSI is given as
\begin{equation}
    {P}_{m_1,j}= \frac{1}{2} \sum\limits_{z=1}^2 \mathrm{Q} \left( \sqrt{2\delta_{m1,j,z}} \right),\ \text{where}\  j=sr,\ s1,\ r1,
\end{equation}
where $\delta_{m1,j,z}$ defines the signal-to-interference plus noise ratio (SINR) for $m_1$ symbols in the presence of HWI and ICSI. It is given as \begin{math}
\delta_{m_1,sr,z}= \frac{P_s \psi_z|\tilde{h}_{sr}|^2}{N_0 + 2P_{s} k_{sr}^2|\tilde{h}_{sr}|^2 + 2(k_{sr}^2 + \psi_z)P_s \sigma_\epsilon^2}
\end{math}, \begin{math}
\delta_{m_1,s1,z}= \frac{P_s \psi_z|\tilde{h}_{s1}|^2}{N_0 + 2P_{s} k_{s1}^2|\tilde{h}_{s1}|^2 + 2(k_{s1}^2 + \psi_z)P_s \sigma_\epsilon^2}
\end{math}, \begin{math}
\delta_{m_1,r1,z}= \frac{P_r \psi_z|\tilde{h}_{r1}|^2}{N_0 + 2P_{r} k_{r1}^2|\tilde{h}_{r1}|^2 + 2(k_{r1}^2 + \psi_z)P_r \sigma_\epsilon^2}
\end{math} where $\psi_z=[(\sqrt{\alpha_1}+\sqrt{\alpha_2})^2,(\sqrt{\alpha_1}-\sqrt{\alpha_2})^2]$ is given according to modulation order.

By using the moment generating function (MGF) and alternate form of  $\mathrm{Q}(.)$ function as being in \cite[eq. (5.4)-(5.6)]{simon2008digital}, the derivation of the average BER (ABER) is defined as follows
\begin{equation}
    {P}_{m_1,j}= \frac{1}{4} \sum\limits_{z=1}^2 \left( 1- \sqrt\frac{\bar{\delta}_{m1,j,z}}{1+\bar{\delta}_{m1,j,z}} \right),  
\end{equation}
where $\bar{\delta}_{m_1,j,z}= E[{\delta}_{m_1,j,z}]$ is defined and
$E[.]$ is an Expectation operator. Thus, 
\begin{math}
\bar{\delta}_{m_1,j,z}= \frac{P_s \psi_z \sigma_{\tilde{h}_{j}}^2}{N_0 + 2P_{s} k_{j}^2\sigma_{\tilde{h}_{j}}^2 + 2(k_{j}^2 + \psi_z)P_s \sigma_\epsilon^2}
\end{math} is given.


As explained above, the (10) gives the BER for a P2P NOMA scheme. Thus, it refers to BER of $m_1$ symbols in downlink NOMA (i.e., $P_{m_1,s1}\triangleq P_{e2e,1}^{(NOMA)}$), the BER of $m_1$ symbols in the first phase of CNOMA and CNOMA-WDL (i.e., $P_{m_1,sr}$), and the BER of $m_1$ symbols in the second phase of CNOMA (i.e., $P_{m_1,r1}$).

To detect $m_2$ symbols, each node implements a SIC process. The BER of $m_2$ symbols at each node is obtained as the sum of the correct and erroneous detection of the $m_1$ symbols during the SIC process. Therefore, by also considering the imperfect SIC,  the BER of $m_2$ at relay and u2 is given by
\begin{equation}
    {P}_{m_2,l}= \frac{1}{2} \sum\limits_{v=1}^6 g_v \mathrm{Q} \left( \sqrt{2\delta_{m_2,l,v}} \right),\ \text{where} \ l=sr,\ s2,\ r2,
\end{equation}
where $\mathbf{g}_v=[1,1,-1,1,1,-1]$, $\delta_{m_2,sr,v}= \frac{P_s \zeta_v|\tilde{h}_{sr}|^2}{N_0 + 2P_{s} k_{sr}^2|\tilde{h}_{sr}|^2 + 2(k_{sr}^2 + \xi_v) P_s \sigma_\epsilon^2}$, $ \delta_{m_2,s2,v}= \frac{P_s \zeta_v|\tilde{h}_{s2}|^2}{N_0 + 2P_{s} k_{s2}^2|\tilde{h}_{s2}|^2 + 2(k_{s2}^2 + \xi_v) P_s \sigma_\epsilon^2}$,  $\delta_{m_2,r2,v}= \frac{P_r \zeta_v|\tilde{h}_{r2}|^2}{N_0 + 2P_{r} k_{r2}^2|\tilde{h}_{r2}|^2 + 2(k_{r2}^2 + \xi_v) P_r \sigma_\epsilon^2}$, $\boldsymbol{\zeta}_v=[\alpha_2, \alpha_2,(\sqrt{\alpha_1}+\sqrt{\alpha_2})^2, (2\sqrt{\alpha_1}+\sqrt{\alpha_2})^2, (\sqrt{\alpha_1}-\sqrt{\alpha_2})^2,(2\sqrt{\alpha_1}-\sqrt{\alpha_2})^2]$ and $\boldsymbol{\xi}_v=[(\sqrt{\alpha_1}+\sqrt{\alpha_2})^2, (\sqrt{\alpha_1}-\sqrt{\alpha_2})^2,(\sqrt{\alpha_1}+\sqrt{\alpha_2})^2,(\sqrt{\alpha_1}+\sqrt{\alpha_2})^2,(\sqrt{\alpha_1}-\sqrt{\alpha_2})^2,(\sqrt{\alpha_1}-\sqrt{\alpha_2})^2]$. In definition of $\zeta_v$, both correct and erroneous SIC conditions are considered. In the case of erroneous SIC, the erroneous detected $m_1$ symbols are subtracted from the received signal, thus the decision rule for detection is shifted. Hence, a few terms in $\zeta_v$ have the coefficient of $2$ which occurs due to the  ipSIC. Besides, one can motice that the priori probability for some terms has a negative value in the coefficient $g_v$ which occurs due to conditional probability in case of ipSIC.   

By using the MGF and  alternate form of $\mathrm{Q}(.)$ function, the ABER of $m_2$ is expressed
\begin{equation}
    {P}_{m_2,l}= \frac{1}{4} \sum\limits_{v=1}^6 g_v\left( 1- \sqrt\frac{\bar{\delta}_{m_2,l,v}}{1+\bar{\delta}_{m_2,l,v}} \right), \end{equation}
where $\bar{\delta}_{m_2,j,v}= E[{\delta}_{m_2,j,v}]$, $\bar{\delta}_{m_2,j,v}= \frac{P_s \zeta_v \sigma_{\tilde{h}_{j}}^2}{N_0 + 2P_{s} k_{j}^2\sigma_{\tilde{h}_{j}}^2 + 2(k_{j}^2 + \xi_v) P_s \sigma_\epsilon^2}$.




As discussed for u1, the BER in (12) gives the error probability of NOMA for a P2P communication. Therefore, it refers to BER of $m_2$ symbols in downlink NOMA i.e., $P_{m_2,s2}\triangleq P_{e2e,2}^{(NOMA)}$), the BER of $m_2$ symbols in the first phase of CNOMA and CNOMA-WDL (i.e., $P_{m_2,sr}$), and the BER of $m_2$ symbols in the second phase of CNOMA (i.e., $P_{m_2,r2}$).

\subsection{BER of cooperative diversity with MRC}
1) At u1: Based on the detected signals in the first phase, R implements a new SC signal again and sends it to u1 and u2 by its power. Thus, u1 and u2 receive two signals from different sources, one from S and one from the R. To improve the users' messages, the users perform an MRC to combine the received signals from the two different phases. The BER of the u1 with diversity combining of the two branches using MRC is given by \cite{kara2018error}, \cite[pp. 320-321]{simon2008digital} 
\begin{equation}
P_{m_1,coop} = \mathrm{Q} \left(\sqrt{2({\delta}_{m_1,s1,z} + {\delta}_{m_1,r1,z})}\right).
\end{equation}
The ABER for cooperative MRC of two branches at u1 when the mean of SNRs of branches are different (i.e.,
\begin{math}
\bar{{\delta}}_{m1,s1,z} \ne \bar{{\delta}}_{m1,r1,z}
\end{math}) is given as in (14) (see the top of the next page).
\begin{figure*}
\begin{equation}
P_{m_1,coop} = \frac{1}{2} \left( 1-\frac{1}{\bar{{\delta}}_{m_1,s1,z}-\bar{{\delta}}_{m_1,r1,z}}\left( \bar{{\delta}}_{m_1,s1,z} \sqrt{\frac{\bar{{\delta}}_{m_1,s1,z}}{1+\bar{{\delta}}_{m_1,s1,z}}} - \bar{{\delta}}_{m_1,r1,z} \sqrt{\frac{\bar{{\delta}}_{m_1,r1,z}}{1+\bar{{\delta}}_{m_1,r1,z}}}
\right)
\right). 
\end{equation}
\hrulefill
\end{figure*}

2) At u2: The u2 detects $m_1$ firstly after combining the received signals with MRC. Thereafter, using the SIC detects its symbols $m_2$. The BER for MRC after the SIC to detect u2 symbols is expressed by \cite{kara2018error},\cite[320-321]{simon2008digital}

\begin{equation}
P_{m_2,coop}=\mathrm{Q} \left(\sqrt{2({\delta}_{m_2,s2,v} + {\delta}_{m_2,r2,v})}\right).
\end{equation}
The ABER for cooperative MRC of two branches at u2 when the mean of the SNRs of branches are different (i.e.,
\begin{math}
\bar{{\delta}}_{m_2,s2,v} \ne \bar{{\delta}}_{m_2,r2,v}
\end{math}) is expressed by \cite{kara2018error}, \cite[pp. 846-847]{JohnG.Proakis2007} as in (16)  (see the top of the next page).
\begin{figure*}
\begin{equation}
P_{m_2,coop} \{s_{sc} \in v\}= \frac{1}{2} \left( 1-\frac{1}{\bar{{\delta}}_{m_2,s2,v}-\bar{{\delta}}_{m_2,r2,v}}\left( \bar{{\delta}}_{m_2,s2,v} \sqrt{\frac{\bar{{\delta}}_{m_2,s2,v}}{1+\bar{{\delta}}_{m_2,s2,v}}} - \bar{{\delta}}_{m_2,r2,v} \sqrt{\frac{\bar{{\delta}}_{m_2,r2,v}}{1+\bar{{\delta}}_{m_2,r2,v}}}
\right)
\right). 
\end{equation}
\hrulefill
\end{figure*}

\subsection{BER of cooperative MRC with error propagation }
Unless a genie-aided relaying is implemented, the R node forwards the detected signals to the users. By considering the ipSIC, the relay can also detect symbols erroneously. Therefore, the erroneous symbols in R are also forwarded to the users. This phenomenon is called error propagation. These erroneous symbols are also combined with the signals received from the direct path. After the implantation of the MRC, we should describe the total received signal in order to acquire the BER in case of error propagation from u1 and u2. Without loosing the generality, we assume that the BS sent a symbol (+1) and a symbol (+1) for u1 and u2, respectively. During the SIC process in R, the symbol of u1 is detected erroneously as (-1). Thus, after the SIC process, the symbol of the u2 is detected erroneously also as (-1). 
\subsubsection{At u1 }
Consequently, the sum of the received signals at the u1 by MRC is given as
\begin{equation}
\begin{split}
& {\varphi}_{m_1,coop}=\tilde{h}^*_{s1} y_1+\tilde{h}_{r1}^* y_{r1} \\&
=({\varphi}_{m_1,s1,z}-{\varphi}_{m_1,r1,z})+\tilde{n}_a,  
\end{split}
\end{equation}
where
\begin{math}
{\varphi}_{m_1,s1,z}= P_s\psi_z |\tilde{h}_{s1}|^2
\end{math} and we define 
\begin{math}
{\varphi}_{m_1,r1,z}= P_r\psi_z |\tilde{g}_{r1}|^2
\end{math}. However, $\tilde{n}_a$ is the effective noise with
$
\mathbf{E}[\tilde{n}_a]\sim \mathcal{CN}(0,\frac{N_0}{2}[\sigma_{\tilde{h}_{s1}}^2 +\sigma_{\tilde{h}_{r1}}^2]+k_{s1}^2 P_s \sigma_{\tilde{h}_{s1}^2} {\sigma_{\epsilon}^2}+k_{s1}^2
P_s\sigma_{\tilde{h}_{s1}}^2+ k_{r1}^2P_r$ \\
$ \sigma_{\tilde{h}_{r1}}^2 {\sigma_{\epsilon}}^2 + k_{r1}^2 P_r \sigma_{\tilde{h}_{r1}}^2+\tau_{m_1,s1,z}\sigma_{\tilde{h}_{s1}}^2\sigma_{\epsilon}^2+\tau_{m_1,r1,z}\sigma_{\tilde{h}_{r1}}^2{\sigma_{\epsilon}^2})$, where 
$\tau_{m_1,s1,z}=P_s\psi_z$
and
$\tau_{m_1,r1,z}=P_r\psi_z$.
Through the MLD decision rule at the u1, $m_1=+1$   is declared if ${\varphi}_{m1,coop} \ge 0$. Thus, the BER of the error propagation for the u1 symbols is defined as
\begin{equation}
\begin{split}
P_{m_1,prop}&= P({\varphi}_{m_1,s1,z}-{\varphi}_{m_1,r1,z}<\tilde{n}_a) \\ &= \mathrm{Q} \left ( \frac{\varphi_{m_1,s1,z}-{\varphi}_{m_1,r1,z}}{\sqrt{\omega_a}} \right),
\end{split}
\end{equation}
where $\omega_a= \frac{N_0}{2}[\sigma_{\tilde{h}_{s1}}^2 +\sigma_{\tilde{h}_{r1}}^2]+k_{s1}^2 P_s {\sigma_{\tilde{h}_{s1}}^2} {\sigma_{\epsilon}^2}+k_{s1}^2 P_s\sigma_{\tilde{h}_{s1}}^2+k_{r1}^2$\\ $P_r\sigma_{\tilde{h}_{r1}}^2 {\sigma_{\epsilon}}^2 +k_{r1}^2P_r \sigma_{\tilde{h}_{r1}}^2+\tau_{m_1,s1,z}\sigma_{\tilde{h}_{s1}}^2\sigma_{\epsilon}^2+\tau_{m_1,r1,z}\sigma_{\tilde{h}_{r1}}^2 {\sigma_{\epsilon}^2}$.
From (18), if the relay forwards an incorrect symbol, this has a significant impact instead of noise on the decision variable MLD.  According to \cite{kara2018error}, we approximate the BER as $\varphi_{m_1,s1,z}-\varphi_{m_1,r1,z}<0$. The BER under error propagation is given as
\begin{equation}
P_{m_1,prop}= P({\varphi}_{m_1,s1,z}-{\varphi}_{m_1,r1,z}<0).
\end{equation}
By averaging (19) over ${\varphi}_{m_1,s1,z}$and ${\varphi}_{m_1,r1,z}$, the average error probability under error propagation is obtained as
\begin{equation}
P_{m_1,prop}= \frac{{\bar\varphi}_{m_1,r1,z}}{{\bar\varphi}_{m_1,r1,z}+{\bar\varphi}_{m_1,s1,z}},
\end{equation} 
where 
\begin{math}
{\bar\varphi}_{m_1,s1,z}=E[
{\varphi}_{m_1,s1,z}]= P_s\psi_z \sigma_{\tilde{h}_{s1}}^2
\end{math} and 
\begin{math}
{\bar\varphi}_{m_1,r1,z}=E[
{\varphi}_{m_1,r1,z}]= P_r\psi_z \sigma_{\tilde{h}_{r1}}^2
\end{math}.

\subsubsection{At u2}
Also, after combining the S and R signals at the u2, the total received signal by u2 is given as
\begin{equation}
{\varphi}_{m_2,coop}=\tilde{h}^*_{s2} y_2+\tilde{h}_{r2}^* y_{r2}.
\end{equation}
After the implementation of the MRC in u2, it detects the u1 symbols first, then it detects its symbols through the SIC process. Therefore, the sum of the received signal at u2 can be indicated as
\begin{equation}
{\varphi}_{m_2,coop}=({\varphi}_{m_2,s2,v}-{\varphi}_{m_2,r2,v})+\tilde{n}_b, 
\end{equation}
where 
${\varphi}_{m_2,s2,v}= P_s\zeta_v |\tilde{h}_{s2}|^2$
 and
${\varphi}_{m_2,r2,v}= P_r\zeta_v |\tilde{h}_{r2}|^2$. $\tilde{n}_b $ is the effective noise with
$\mathbf{E}[\tilde{n}_b] \sim\mathcal{CN}(0,\frac{N_0}{2}[\sigma_{\tilde{h}_{s2}}^2 +\sigma_{\tilde{h}_{r2}}^2]+k_{s2}^2 P_s {\sigma_{\tilde{h}_{s2}}^2} {\sigma_{\epsilon}^2}+k_{s2}^2
P_s\sigma_{\tilde{h}_{s2}}^2 +k_{r2}^2 P_r$, $\sigma_{\tilde{h}_{r2}}^2 {\sigma_{\epsilon}}^2+ k_{r2}^2 P_r \sigma_{\tilde{h}_{r2}}^2+\tau_{m_2,s2,v}\sigma_{\tilde{h}_{s2}}^2 \sigma_{\epsilon}^2+\tau_{m_2,r2,v}\sigma_{\tilde{h}_{r2}}^2 {\sigma_{\epsilon}^2})
$, where
${\tau}_{m_2,s2,v}= P_s\zeta_v $,
and
${\tau}_{m_2,r2,v}= P_r\zeta_v $.
Through the MLD decision rule and the SIC process at the u2 are declared as $m_1=+1$ and $m_2=+1$. Thus, the BER of the error propagation for the u2 symbols is defined as
\begin{equation}
\begin{split}
P_{m_2,prop}&= P({\varphi}_{m_2,s2,v}-{\varphi}_{m_2,r2,z}<\tilde{n}_b) \\ & = \mathrm{Q} \left ( \frac{\varphi_{m_2,s2,v}-{\varphi}_{m_2,r2,v}}{\sqrt{\omega_b}} \right),
\end{split}
\end{equation}
where $\omega_b= \frac{N_0}{2}[\sigma_{\tilde{h}_{s2}}^2 +\sigma_{\tilde{h}_{r2}^2}+k_{s2}^2 P_s {\sigma_{\tilde{h}_{s2}}^2} {\sigma_{\epsilon}^2}+k_{s2}^2 P_s\sigma_{\tilde{h}_{s2}}^2 + $
$k_{r2}^2 P_r \sigma_{\tilde{h}_{r2}}^2 {\sigma_{\epsilon}}^2 + k_{r2}^2 P_r \sigma_{\tilde{h}_{r2}}^2+\tau_{m_2,s2,v}\sigma_{\tilde{h}_{s2}}^2\sigma_{\epsilon}^2+\tau_{m_2,r2,v}\sigma_{\tilde{h}_{r2}}^2 {\sigma_{\epsilon}^2}$
From (23), if R forwards an incorrect symbol, this has a significant impact instead of noise on the decision variable MLD and SIC. Consequently, we approximate the BER for the SIC to detect $m_2$ as$ {\varphi_{m_2,s2,v}-{\varphi}_{m_2,r2,v}}<0$\cite{kara2018error}, the BER under error propagation is presented as
\begin{equation}
P_{m_2,prop}= P({\varphi}_{m_2,s2,v}-{\varphi}_{m_2,r2,v}<0).
\end{equation}
By averaging (24) over ${\varphi}_{m_2,s2,v}$ and ${\varphi}_{m_2,r2,v}$ , the average error probability under error propagation is obtained as
\begin{equation}
P_{m_2,prop}= \frac{{\bar\varphi}_{m_2,r2,v}}{{\bar\varphi}_{m_2,r2,v}+{\bar\varphi}_{m_2,s2,v}}.
\end{equation}
where
${\bar\varphi}_{m_2,s2,v}=E[{\varphi}_{m_2,s2,v}]= P_s\zeta_v \sigma_{\tilde{h}_{s2}}^2$
 and
${\bar\varphi}_{m_2,s2,v}=E[{\varphi}_{m_2,r2,v}]= P_r\zeta_v \sigma_{\tilde{h}_{r2}}^2$.

Finally, to find the e2e BER in CNOMA, we obtain $P_{m_1,sr}$, $P_{m_1,r1}$ by using (10) and $P_{m_2,sr}$, $P_{m_1,r2}$ by using (12) and  substitute them into (7) for u1 and u2, respectively. On the other hand, to obtain the e2e BER in CNOMA-WDL, we determine $P_{m_1,sr}$, $P_{m_1,coop}$ and $P_{m_1,prop}$ by using (10), (14) and (20) and substitute them into (8) for u1. Similarly, we determine $P_{m_2,sr}$, $P_{m_2,coop}$ and $P_{m_2,prop}$ by using (12), (16) and (25) and substitute them into (8) for u2.

\section{Numerical results}
In this section, we validate the analytical BER results with computer simulations for three schemes (i.e., NOMA, CNOMA, CNOMA-WDL). Unless otherwise stated, we set the parameters to $d_{s1}=4m$, $d_{s2}=2m$, $d_{sr}=1m$, $d_{r1}=3m$, $d_{r2}=1m$,  $P_r=P_s$, $a=2$, $\alpha_1=0.8$, $\alpha_2=0.2$ and the HWI level is equal for all nodes, i.e., $k_{s1}=k_{s2}=k_{r1}=k_{r2}=k$.

In Fig. 2.a, we present the BER performance w.r.t SNR where the ICSI is considered as $\sigma_\epsilon^2=0.005$ and the HWI factors are set at $k=0.175$ representing the worst value of HWI in open literature. First, it is observed that the numerical results match perfectly with the simulation results, which proves the correctness of our analysis. Based on the results, as expected, with the increase of HWI and ICSI levels , all NOMA schemes get worse performance, which reflect to practical implementations. By comparing NOMA schemes with the HWI, we obverse that NOMA is superior to CNOMA. This can be explained as follows. In the presence of HWI, due to the increased HWI in total (additional HWI at R node), the performance of each phase is drastically degraded, so that the e2e performance of CNOMA becomes worse than simple downlink NOMA schemes. On the other hand, CNOMA-WDL outperforms both schemes whether a HWI is introduced or not since a diversity path is achieved by MRC. Nevertheless, a full diversity order (i.e., $2$) may not be observed. This is for two reasons. The first is due to error propagation from relay to users. As explained in analysis, unless a genie-aided relaying is not considered, the relay node also forwards erroneous symbols to the users, which causes an error floor in high SNR regime. The second reason of the error floor is the imperfections in the system due to  HWI or ICSI.
\begin{figure*}
\centering
\subfloat{\includegraphics[width=.66\columnwidth]{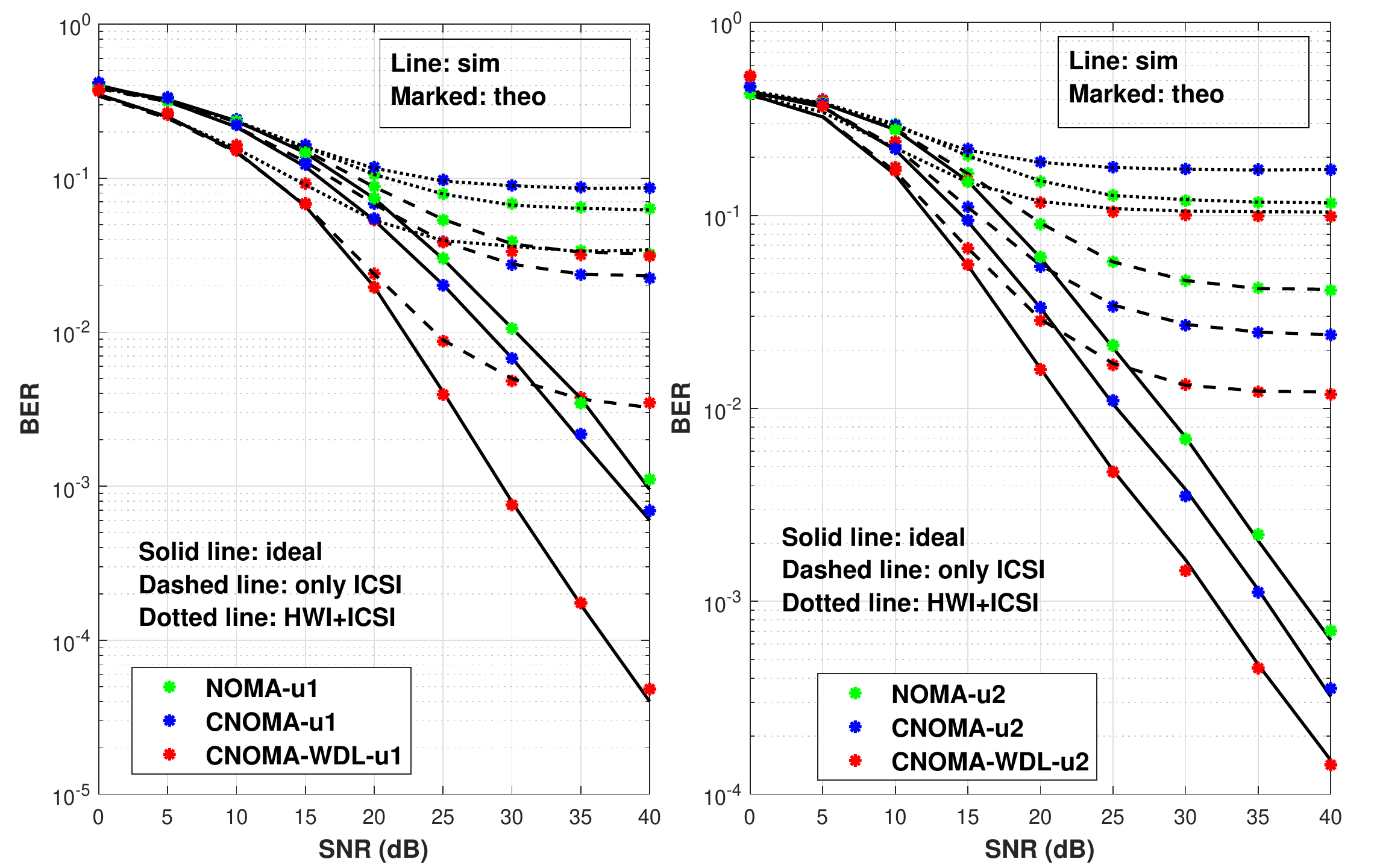}\label{subf1}}
\subfloat{\includegraphics[width=.66\columnwidth]{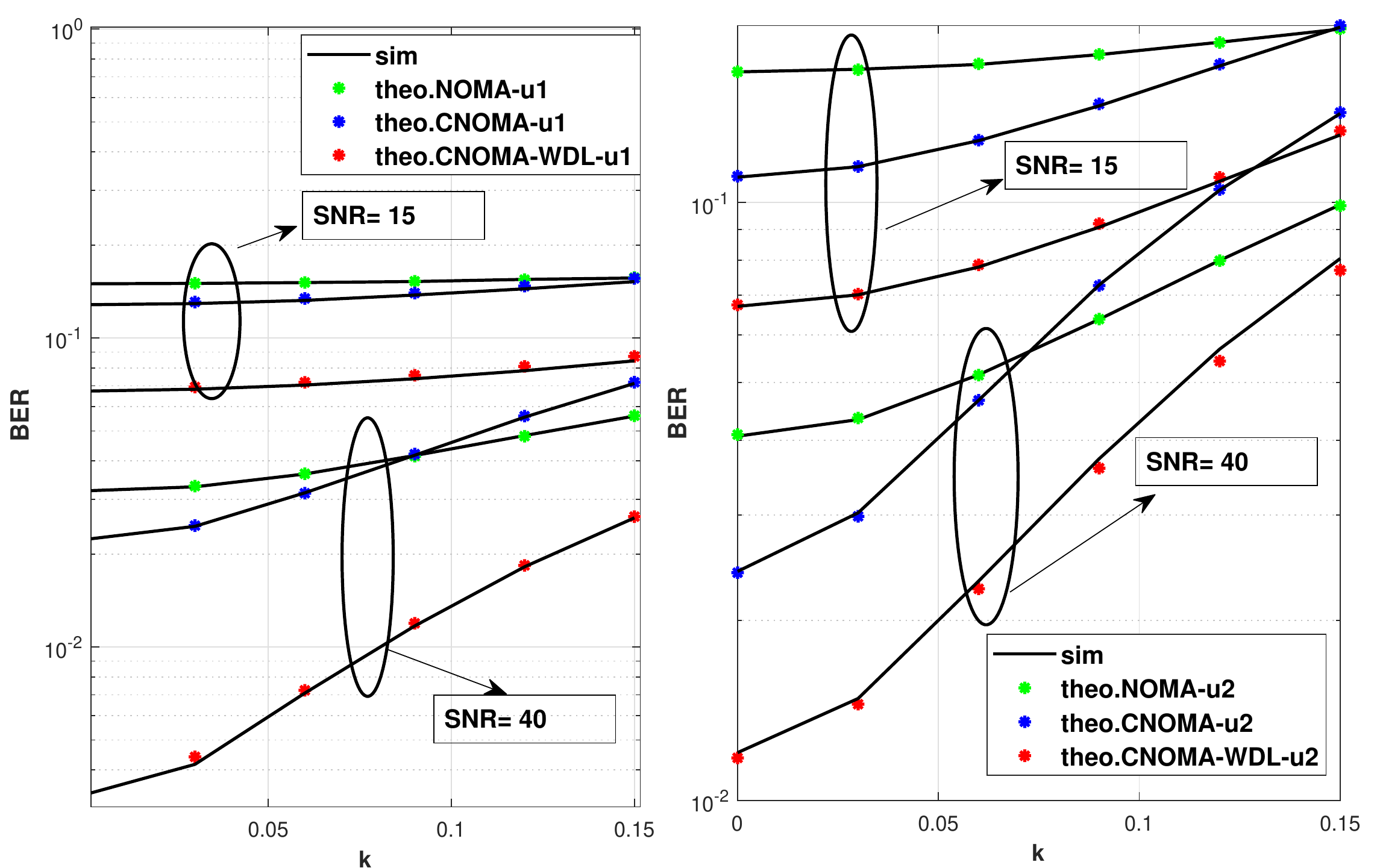}\label{subf2}\label{subf3}}
\subfloat{\includegraphics[width=.66\columnwidth]{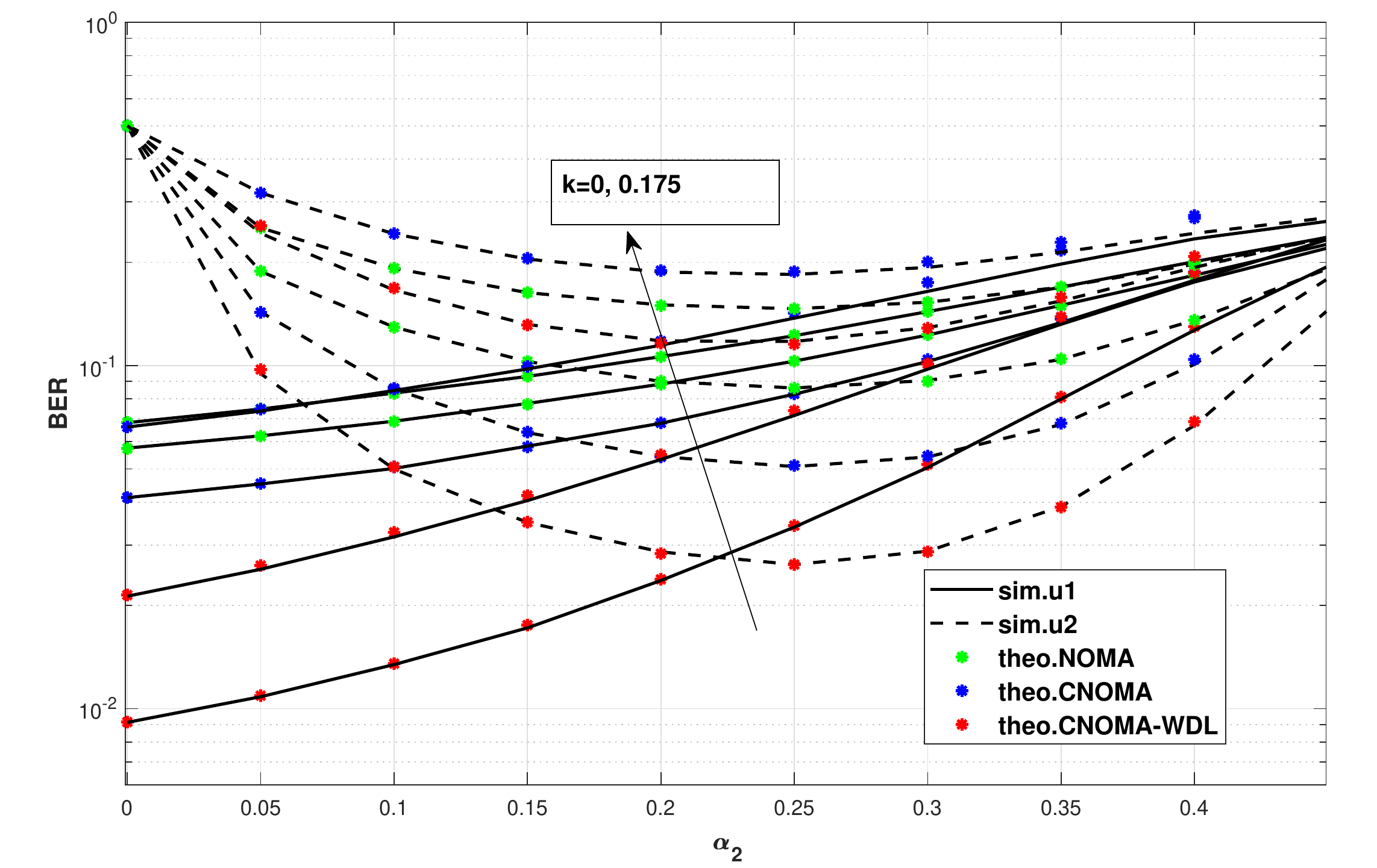}}
\caption{BER performances of NOMA, CNOMA and CNOMA-WDL a) w.r.t SNR when  $\sigma_\epsilon^2= 0.005$ b)  w.r.t HWI level when  $\sigma_\epsilon^2= 0.005$ and $SNR=15,\ 40$ dB c) w.r.t PA  when  $\sigma_\epsilon^2= 0.005$ and $SNR=20$ dB. }
    \label{constellatio}
\end{figure*}


To further evaluate the effect of HWI, in Fig. 2.b, we present BER performances versus the HWI level for $SNR=15,\ 40$ dB when $\sigma_\epsilon^2=0.005$. As expected, with increasing HWI, all schemes have a higher BER. In Fig. 2.b, one can easily see that the CNOMA-WDL is superior to both schemes for all HWI values. For $SNR=15$ dB, the CNOMA outperforms the NOMA regardless of HWI level. However, for $SNR=15$ dB, NOMA becomes better than CNOMA after $k\approx 0.1$ since, as explained in the previous figure, with increasing HWI, an error floor occurs for both phases (i.e., $S-R$ and $R-Ui, \ i=1,2$). Hence, the performance of CNOMA decreases dramatically, so NOMA provides better performance. In this regard, we can say that when the HWI is high, using NOMA rather than CNOMA is more beneficial.
    
Fig. 2.c shows the effect of power allocation (PA) on the BER performances with different levels of HWI. It is observed that the PA affects the BER performance of one user at the expense of the other. Nevertheless, increasing $\alpha_2$ too much does always not mean an increase in performance of u2 due to SIC process. Thus, PA should be carefully chosen not to cause an unfairness for users.


\section{Conclusion}
In this letter, we investigate the BER performance of three NOMA schemes (i.e., downlink NOMA, cooperative NOMA with and without direct links) under practical assumptions. We consider all imperfections (i.e., ipSIC, ICSI, HWI) and derive the exact e2e BER expressions for all schemes. All theoretical deductions are validated with computer simulations. We reveal that NOMA can outperform CNOMA when the HWI is relatively high. Besides, we discuss the effect of PA on the BER performances and present that the PA  has an important role; thus, it should be chosen wisely no to cause an error propagation whether from relay to users in CNOMA-WDL or an error in SIC processes in all schemes. We believe that this letter can provide fruitful insights for the practical implementation of NOMA schemes and the analysis in this work help researchers to make further analysis for other NOMA schemes under practical assumptions.
\bibliographystyle{IEEEtran}
\bibliography{reference}

\begin{thebibliography}{10}
\providecommand{\url}[1]{#1}
\csname url@samestyle\endcsname
\providecommand{\newblock}{\relax}
\providecommand{\bibinfo}[2]{#2}
\providecommand{\BIBentrySTDinterwordspacing}{\spaceskip=0pt\relax}
\providecommand{\BIBentryALTinterwordstretchfactor}{4}
\providecommand{\BIBentryALTinterwordspacing}{\spaceskip=\fontdimen2\font plus
\BIBentryALTinterwordstretchfactor\fontdimen3\font minus
  \fontdimen4\font\relax}
\providecommand{\BIBforeignlanguage}[2]{{%
\expandafter\ifx\csname l@#1\endcsname\relax
\typeout{** WARNING: IEEEtran.bst: No hyphenation pattern has been}%
\typeout{** loaded for the language `#1'. Using the pattern for}%
\typeout{** the default language instead.}%
\else
\language=\csname l@#1\endcsname
\fi
#2}}
\providecommand{\BIBdecl}{\relax}
\BIBdecl

\bibitem{zeng2020cooperative}
M.~Zeng, W.~Hao, O.~A. Dobre, and Z.~Ding, ``{Cooperative NOMA: State of the
  art, key techniques, and open challenges},'' \emph{IEEE Netw.}, vol.~34,
  no.~5, pp. 205--211, 2020.

\bibitem{Liu2018}
H.~Liu, Z.~Ding, K.~J. Kim, K.~S. Kwak, and H.~V. Poor, ``{Decode-and-forward
  relaying for cooperative NOMA systems with direct links},'' \emph{IEEE Trans.
  Wirel. Commun.}, vol.~17, no.~12, pp. 8077--8093, 2018.

\bibitem{smaini2012rf}
L.~Smaini, \emph{{RF Analog Impairments Modeling for Communication systems
  Simulation: Application to OFDM-based Transceivers}}.\hskip 1em plus 0.5em
  minus 0.4em\relax John Wiley $\&$ Sons, 2012.

\bibitem{li2019effects}
M.~Li, B.~Selim, S.~Muhaidat, P.~C. Sofotasios, M.~Dianati, P.~D. Yoo,
  J.~Liang, and A.~Wang, ``{Effects of residual hardware impairments on secure
  NOMA-based cooperative systems},'' \emph{IEEE Access}, vol.~8, pp.
  2524--2536, 2019.

\bibitem{deng2020hardware}
C.~Deng, M.~Liu, X.~Li, and Y.~Liu, ``{Hardware impairments aware full-duplex
  NOMA networks over Rician fading channels},'' \emph{IEEE Syst. J.}, vol.~15,
  no.~2, pp. 2515--2518, 2020.

\bibitem{li2019residual}
X.~Li, J.~Li, Y.~Liu, Z.~Ding, and A.~Nallanathan, ``{Residual transceiver
  hardware impairments on cooperative NOMA networks},'' \emph{IEEE Trans.
  Wirel. Commun.}, vol.~19, no.~1, pp. 680--695, 2019.

\bibitem{arzykulov2021hardware}
S.~Arzykulov, G.~Nauryzbayev, A.~Celik, and A.~M. Eltawil, ``{Hardware and
  interference limited cooperative CR-NOMA networks under imperfect SIC and
  CSI},'' \emph{IEEE Open J. Commun. Soc.}, vol.~2, pp. 1473--1485, 2021.

\bibitem{hamza2021error}
A.~A. Hamza, I.~Dayoub, I.~Alouani, and A.~Amrouche, ``{On the error rate
  performance of full-duplex cooperative NOMA in wireless networks},''
  \emph{IEEE Trans. Commun.}, vol.~70, no.~3, pp. 1742--1758, 2022.

\bibitem{khennoufa2022bit}
F.~Khennoufa, K.~Abdellatif, and F.~Kara, ``{Bit error rate and outage
  probability analysis for multi-hop decode-and-forward relay-aided NOMA with
  imperfect SIC and imperfect CSI},'' \emph{AEU-Internat. J. Electron.
  Commun.}, vol. 147, p. 154124, 2022.

\bibitem{mohjazi2019error}
L.~Mohjazi, L.~Bariah, S.~Muhaidat, P.~C. Sofotasios, O.~Onireti, and M.~A.
  Imran, ``{Error probability analysis of non-orthogonal multiple access for
  relaying networks with residual hardware impairments},'' in \emph{IEEE 30th
  Annual Internat. Symp. Pers., Indoor Mobile Radio Commun. (PIMRC)}, 2019.

\bibitem{afana2018joint}
A.~Afana, N.~Abu-Ali, and S.~Ikki, ``{On the joint impact of hardware and
  channel imperfections on cognitive spatial modulation MIMO systems:
  Cramer--Rao bound approach},'' \emph{IEEE Syst. J.}, vol.~13, no.~2, pp.
  1250--1261, 2018.

\bibitem{kara2018error}
F.~Kara and H.~Kaya, ``{On the error performance of cooperative-NOMA with
  statistical CSIT},'' \emph{IEEE Commun. Lett.}, vol.~23, no.~1, pp. 128--131,
  2018.

\bibitem{simon2008digital}
M.~K. Simon and M.-S. Alouini, \emph{{Digital Communications over Fading
  Channels }}.\hskip 1em plus 0.5em minus 0.4em\relax John-Wiley $\&$ Sons,
  2004.

\bibitem{JohnG.Proakis2007}
J.~G. Proakis and M.~Salehi, \emph{{Digital Communications}}.\hskip 1em plus
  0.5em minus 0.4em\relax McGraw-Hill, 2008.

\end{thebibliography}
\end{document}